# Highly ionized xenon and volumetric weighting in restricted focal geometries

J. Strohaber*, A. A. Kolomenskii and H. A. Schuessler

*Florida A&M University, Department of Physics, Tallahassee, FL 32307, USA*

*Texas A&M University, Department of Physics, College Station, TX 77843-4242, USA*

[*]*Corresponding author: james.strohaber@famu.edu*

**Abstract:** The ionization of xenon atoms subjected to 42 fs, 800nm pulses of radiation from a Ti:Sapphire laser was investigated. In our experiments, a maximum laser intensity of $\sim 2\times 10^{15}\,\text{W}/\text{cm}^2$ was used. Xenon ions were measured using a time-of-flight ion mass spectrometer having an entrance slit with dimensions of $12\,\mu m \times 400\,\mu m$. The observed yields $\text{Xe}^{n+}$ ($n=1-7$) were partially free of spatial averaging. The ion yields showed sequential and nonsequential multiple ionization and dip structures following saturation. To investigate the dip structures and to perform a comparison between experimental and simulated data, with the goal of clarifying the effects of residual spatial averaging, we derived a hybrid analytical-numerical solution for the integration kernel in restricted focal geometries. We simulated xenon ionization using Ammosov-Delone-Krainov and Perelomov-Popov-Terent'ev theories and obtained agreement with the results of observations. Since a large number of experiments suffer from spatial averaging, the results presented are important to correctly interpret experimental data by taking into account spatial averaging.

## I. INTRODUCTION

Over the past few decades, research involving the interaction of intense and ultrashort pulsed radiation with matter has become common place and made possible by the advent of chirped-pulse amplification (CPA) [1]. Pulses of radiation generated in a CPA laser have a number of useful properties such as high achievable peak intensities, which provide the high optical pumping rates necessary for multiphoton processes as described by lowest order perturbation theory (LOPT) [2], a broadband spectrum having applications in spectroscopy [3], and the production of short pulses used in pump- probe experiments to investigate ultrafast phenomena on the time scale of molecular vibrations [4] and on the time scale of electron dynamics [5]. In the quantitative analysis of the products of laser-matter interactions, a versatile instrument known as a time-of-flight ion mass spectrometer can be used [6]. In these types of experiments, an intense laser beam is focused into a vacuum chamber where it is allowed to inter- act with chosen target particles. Product ions can then be directed towards a detector such as a multichannel plate (MCP), delay line detector, or Faraday cup for quantification using ion optics. It was known early on that the production of ions using laser beams resulted in an averaging effect that leads to an $I^{3/2}$ dependence in measured yield curves [7]. This unwanted experimental artifact manifests itself in measuring incorrect relative ion yields between charge states and fragment ions [12], and averaging over structures in intensity dependent ionization yields [9—11]. This process is known in the literature as spatial or intensity averaging, or volumetric weighting [12]. Because of spatial averaging, one does not measure the ionization probability but an averaged result. For researchers in this area, spatial averaging frustrates comparison of experimental data with theoretical results, and many instances can be found in the literature, where theoretical results are artificially averaged for comparison [13, 14]. Averaging probabilities tend to make all intensity-dependent ion yield curves similar, and interpretation becomes generic. With the exception of dominating processes such as the nonsequential double ionization of helium, only characteristics such as the order of the multiphoton process and saturation intensity may be determined, while the remaining photophysical phenomena are masked. The potential benefits of intensity-resolved measurements have fueled the development of a plethora of techniques designed specifically for unraveling the masking effects of spatial averaging. These methods can be divided into three groups: pure theoretical, [10, 15] pure experimental, [6, 16] and combined approaches [8,12,17,18]. Pure theoretical approaches involve mathematical algorithms used to deconvolve experimental data; for instance, researchers working with ion beams typically employ the Abel transformation, which allows the Newton sphere to be retrieved. More recently, deconvolution of photoelectron yields, in which all electrons in the focus were collected, was performed by a variational approach. As of yet, no purely experimental method has been developed to measure ionization probabilities in above threshold ionization (ATI). However, considering ion detection currently, two purely experimental techniques exist that have demonstrated successful results. The first is the photo- dynamic test tube pioneered by Strohaber and Uiterwaal6 and the second is the ion microscope [16]. Finally, mixed methods include intensity-selective scanning (ISS)8,12 and intensity difference scanning (IDS) [17,18]. With the exception of the pure theoretical approach, many experiments rely on some type of aperture for data collection. This work will provide insight into the effects of volumetric weighting on data, and a complete understanding of the role that volumetric weighting plays in the interaction of radiation with matter in experiments having restricted focal geometry.

Due to the large number of possible experimental configurations, their effects on volumetric weighting of data will be considered. Finally, a method with the intent to compare data measured for different experimental configurations is developed. This analysis is largely fueled by the observation of anomalous structures in measured xenon yields. It has three immediate consequences for research in ionization in strong fields: (i) the ability to correctly interpret experimental data, (ii) to properly spatially average theoretical ionization probabilities for comparison with measured data, and (iii) to deconvolve measured data obtained from various focal and slit geometries.

## II. EXPERIMENTAL RESULTS ON THE IONIZATION OF XENON

In our ionization experiments with xenon, ultrashort pulses (42 fs) were produces by a Spectra-Physics Spitfire CPA laser having a repetition rate of 1 kHz. The radiation had a center wavelength of 800 nm and a pulse energy of 2.5 mJ. Radiation was focused into an ionization chamber by a lens having a nominal focal length of 21 cm. The focused radiation had a minimum beam size of $w_0 = 35.5 \mu m$ at the $1/e^2$ level and a Rayleigh range of $z_0 \approx 4.9$ mm. A detailed description of the time-of-flight apparatus is given in Refs. [6] and [9]. In brief, ions are created within a parallel plate capacitor through photoionization. The produced ions are accelerated into a flight tube where they are subsequently detected by a MCP. By using a combination of spatial (slit) and temporal (time slicing) filtering, ions can be selectively detected from a variable but limited three-dimensional spatial region within the focus.

The backing pressures of both the ionization chamber and flight tube were $\sim 5 \times 10^{-9}$ mbar. Xenon gas having a purity of 99.999% was admitted into the ionization chamber using a precision leak valve (MDC, ULV-150) to a pressure of $\sim 2 \times 10^{-7}$ mbar. Data were recorded using a FAST ComTec counting card (P7886 2-GHz) having a time resolution of 0.5 ns. Data acquisition was automated with LabVIEW code, which measured and adjusted the laser power and recorded experimental parameters and the spectrum. The laser power was attenuated by adjusting a half waveplate positioned before the compressor. For each laser power, the spectrum was averaged over 150,000 laser pulses resulting in a total maximum runtime of 5 h. The laser powers used in our experiments corresponded to intensities ranging between $4.7 \times 10^{13}$ W/cm$^2$ and $2.3 \times 10^{15}$ W/cm$^2$. Figure 1 shows the intensity-dependent ionization yields for xenon ions Xe$^{n+}$ ($n = 1-7$) measured with the time of flight apparatus. The data were taken using a slit with dimensions of 400 $\mu$m along the Rayleigh range and 12 $\mu$m in the transverse direction. The remaining dimension of the detection volume was unrestricted (no time slicing).

The data shown in Fig. 1 are not only indicative of a sequential ionization process but also show contributions from nonsequential multiple ionization. The classical description of over the barrier ionization (OBTI) can be used to estimate the saturation intensities Is as the intensity required to lower the field-induced barrier to the energy of the ground state $I_{OBTI} = 4 \times 10^9 (IE^4 / Z^2)$. Here, IE is the ionization energy and Z is the final charge state. Using IEs of [12.13, 20.98, 31.05, 42.20, 54.10, 66.70] eV, obtained from the NIST database [19], the saturation intensities are found to be [0.87, 1.94, 4.13, 7.93, 13.71, 22.00] $\times 10^{14}$ W/cm$^2$. These values are shown by vertical dotted lines in Fig. 1(b). Around $\sim 10^{14}$ W/cm$^2$, a shoulder structure can be seen in the Xe$^{2+}$ ion yields. This shoulder structure saturates near the expected saturation intensity of the respective previous charge state Xe$^{1+}$. Other charge states show indications of nonsequential processes, but we have pointed out the most noticeable one. The data also show a new structure that has not been previously observed. Following saturation of the ion yields of charge states 2, 3, and 4, dips appears as shown in Fig. 1(b). These dips occur near the saturation of the following ionic state. It is suspected that the dips are, therefore, formed by the higher nonlinear processes of the subsequent charge states followed by an increase due to spatial averaging and may indicate that the dips are due to a genuine physical process. Since, in general, reducing the effects of spatial averaging (i.e., by restricting the focal volume) produces data that are expected to be more representative of true ionization probabilities, further investigation is required.

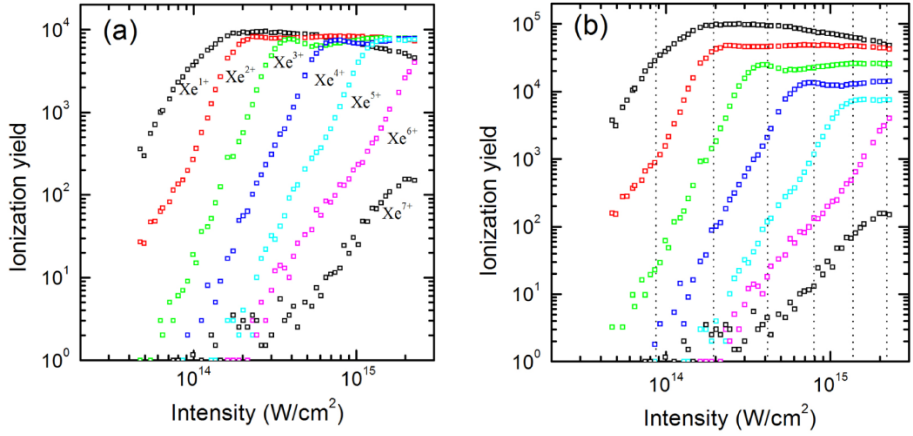

FIG. 1. Ionization yields of xenon using 42 fs pulses of radiations centered at 800 nm. (a) The yield curves are those of $Xe^{n+}$, where $n = 1-7$. Charges states show clear evidence of nonsequential multiple ionization. Dips in the yield curves following saturation can be seen for charge states 2, 3, and 4. (b) Same as (a) except yields have been vertically displaced by $\times 2$, $\times 4$, $\times 8$, and $\times 10$ for $Xe^{4+}$, $Xe^{3+}$, $Xe^{2+}$, and $Xe^{1+}$, respectively. Vertical dotted lines indicate saturation intensities determined by OBTI, see text.

To investigate the origin of these dips, we simulated ionization yield curves using ionization rates calculated with the Ammosov-Delone-Krainov (ADK) and Perelomov-Popov-Terent'ev (PPT) theories [20, 21]. While the data in Fig. 1 are to a large degree free of spatial averaging, there remains a residual amount; therefore, to compare with experimental data, the theoretical results must be spatially averaged with the volumetric weighting of the detection volume used in the experiment. Previously, an attempt to approximate the kernel [22] for a restricted focal volume was made; however, the results were at best qualitative and thus not suitable for accurate calculations. For these reasons, we have derived a hybrid analytical-numerical solution for the integration kernel and investigated its characteristics and its effects on ionization probabilities. To the best of our knowledge, the kernel for an arbitrary restricted focal geometry has not been presented in the literature and because of its importance in numerous types of experiments, we present results that are general and can be directly applied to a variety of experimental configurations where volumetric weighting is known to obscure results.

## III. VOLUMETRIC WEIGHTING FACTORS

When measuring ionization yields in an experiment, the collected ions originate from different locations in the focus, and because the intensity distribution is not uniform (typically taken to be Gaussian), the collected ions are those produced over a broad range of intensities. The ion yields S(I0) in an experiment at a peak intensity of I0 are the sum of the products of the probability at a local intensity and the volume of the iso-intensity shell at that intensity. This can be written as the integrated product yield

$$S(I_0) \propto \int_0^{I_0} P(I)\left|\frac{\partial V}{\partial I}\right| dI \tag{1}$$

Here, $P(I)$ is the actual ionization probability and is the sought after quantity, and $K(I, I_0) = \partial V / \partial I$ is the volumetric weighting factor or integration kernel, a quantity that depends on both the local and peak intensities. The kernel is described completely by the focal geometry. In this work, the kernel is taken to be that due to the Gaussian intensity profile

$$I(z) = I_0 \frac{w_0^2}{w(z)^2} \exp\left(-\frac{2r^2}{w(z)^2}\right) \tag{2}$$

Here, $w_0$ is the waist, $w^2 = w_0^2(1 + z^2/z_0^2)$ is the spot size, and $z_0$ is the Rayleigh range of the beam. At a specified intensity $I$, the radius of the corresponding isointensity shell as a function of $z$ can be found from Eq. (2)

$$r(z) = w(z)\sqrt{\frac{1}{2}\ln\left|\frac{I_0 w_0^2}{I w^2(z)}\right|} \tag{3}$$

This expression has the restriction that the natural log under the radical must be greater than or equal to zero, which implies that the natural log argument is greater than one, $I_0 w_0^2 \geq I w^2(z)$. This requirement gives the extent of the iso-intensity shells along the $z$-direction $z_\pm = \pm z_0 \sqrt{I_0/I - 1}$. The volume is then found by integration

$$V_{3D} = \iiint dV = \pi \int_{z_-}^{z_+} r^2(z) dz \tag{4}$$

Making use of Eq. (3) and integrating the last expression in Eq. (4) gives

$$V_{3D} = z_0 w_0^2 \frac{\pi}{9}\left[\left(\frac{I_0}{I} - 1\right)^{3/2} + 6\left(\frac{I_0}{I} - 1\right)^{1/2} - 6\arctan\left(\sqrt{\frac{I_0}{I} - 1}\right)\right] \tag{5}$$

Equation (5) is the well-known result for the volume within the iso-intensity shells in a Gaussian focus [23].

By restricting the focal volume along the *z*-direction, usually accomplished by a slit (ISS approach), the volume for this restricted geometry is found by the last integral of Eq. (4) with integration bounds of $z' = z - c/2$ and $z' = z + c/2$,

$$V_{3D}^{RV_z} = z_0 w_0^2 \frac{\pi}{18} \left[ 3\frac{z}{z_0}\left(3 + \frac{z^2}{z_0^2}\right)\ln\left(\frac{I_0}{I}\frac{z_0^2}{z_0^2 + z^2}\right) + 2\frac{z}{z_0}\left(6 + \frac{z^2}{z_0^2}\right) - 12\arctan\left(\frac{z}{z_0}\right) \right]_{z-c/2}^{z+c/2} \quad (6)$$

Here, $c$ is the length of the slit in the *z*-direction. For the given volumes in Eqs. (5) and (6), the associated kernels found by taking the derivative with respect to intensity are

$$K_{2D}^{FV}(I, I_0) = \pi \frac{w^2}{2I} \quad (7a)$$

$$K_{3D}^{RV_z}(I, I_0) = \pi c \frac{w_0^2}{2I}\left(1 + \frac{z^2}{z_0^2} + \frac{c^2}{12 z_0^2}\right) \quad (7b)$$

$$K_{3D}(I, I_0) = \pi z_0 w_0^2 \frac{1}{3I}\sqrt{\frac{I_0}{I} - 1}\left(\frac{I_0}{I} + 2\right) \quad (7c)$$

Equation (7a) is the kernel of a 2D slice of zero thickness taken perpendicular through the beam $V_{2D} = \iint dA = \pi r^2$ and is full-view in 2D. The kernel in Eq. (7b) represents a slice of finite extent taken in the transverse direction and centered at position z. This kernel is different than that given in Refs. [8] and [12]. In Ref. [12], the kernel is entirely due to a 2D configuration, and this small detail has noticeable effects when computing the kernel in a restricted geometry. Equation (7c) is the full-view kernel in 3D.

### IV. VOLUMETRIC WEIGHTING IN RESTRICTED FOCAL GEOMETRIES

In many experiments, a rectangular slit is used to restrict the volume of the focus exposed to the detector. In these cases, the volumetric weighting factors are more complex than those given by the analytical expressions in Eqs. (7a) and (7c). To obtain the kernel associated with a 3D detection volume within the focus, we integrate $dV$ in a restricted space over the bounds of a rectangular parallelepiped detection volume. The sides of our detection volume have lengths of $a$ and $b$ in the transverse dimension, and length $c$ in the longitudinal direction. An analytical solution to this integral in terms of elementary functions does not exist, and for this reason, we have developed a hybrid analytical-numerical solution obtained by adding up transverse slices of exact solutions along the propagation direction. To do this, three different cases must be considered to determine the volume within an iso-intensity shell from a restricted volume. These three cases are illustrated in Fig. 2 with a detection volume of dimensions b(vertical) and a (horizontal).

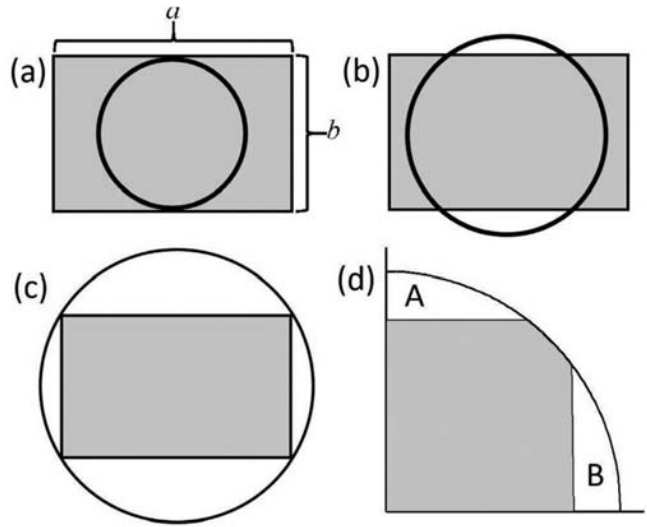

FIG. 2. Two dimensional detection volume and iso-intensity shells. (a) Iso-intensity shell (circle) with diameter smaller than the smallest length of the detection volume (rectangle). (b) Iso-intensity shell with diameter larger than one side of the detection volume (intermediate case). (c) Isointensity shell with a radius larger than half of the diagonal length of the detection volume. (d) Quadrant view of an intermediate case similar to that in (b) with the detected ions coming from the shaded region. The detection volume is the volume in the iso-intensity shell minus parts A and B.

In Fig. 2(a), iso-intensity shells (circle) with $r < b/2$ are without restriction, in Fig. 2(b), the radius of the iso-intensity shell is greater than $b = 2$ but less than $a = 2$, in Fig. 2(c), the detection volume is completely enclosed in the iso-intensity shell. Fig. 2(d) is a quadrant view of an intermediate case similar to that shown in Fig. 2(b). For shells completely enclosed [Fig. 2(a)], the volume within the 2D shells is

$$V_{2D}^{(1)} = \pi r^2, \tag{8}$$

where $r$ is given in Eq. (2). In Fig. 2(c), the shell radius is larger than half the diagonal length $r \geq \sqrt{a^2 + b^2}/2$, and in this case, the volume of the iso-intensity shell is equal to the volume of the detection volume $V_{2D}^{(3)} = ab$. In the case where the radius of the iso-intensity shell is $(a/2 \text{ or } b/2) \leq r \leq \sqrt{a^2 + b^2}/2$, the volume is more complicated; however, an exact analytical expression can be derived. In Fig. 2(d), a quarter of the detection volume (shaded region) is shown. The detection volume is the volume within the iso-intensity shell (circle) minus four times the volume of parts A and B or $V_{2D}^{(2)} = \pi r^2 - 4V_A - 4V_B$. Part B is the region bounded by the curves $y = \sqrt{r^2 - x^2}$ and $y = 0$, and its volume is found by integrating from $x = a/2$ to $x = r$. Similar analysis can be made for region A and results in the following volumes:

$$V_A = \frac{1}{4}\left[\pi r^2 - a\sqrt{r^2 - \frac{a^2}{4}} - 2r^2 \arcsin\left(\frac{a}{2r}\right)\right] \tag{9a}$$

$$V_B = \frac{1}{4}\left[\pi r^2 - b\sqrt{r^2 - \frac{b^2}{4}} - 2r^2 \arcsin\left(\frac{b}{2r}\right)\right] \tag{9b}$$

The total restricted volume is therefore

$$V_{2D}^{(2)} = -\pi r^2 + a\sqrt{r^2 - \frac{a^2}{4}} + b\sqrt{r^2 - \frac{b^2}{4}} + 2r^2\left[\arcsin\left(\frac{a}{2r}\right) + \arcsin\left(\frac{b}{2r}\right)\right] \tag{10}$$

The volumes in Eqs. (8) and (10) together with $V_{2D}^{(3)} = ab$ are plotted (as a function of radius) separately in Fig. 3 as the solid blue (region I), red (region II), and black (region III) curves, respectively. The dimensions used for the volume in Fig. 3(a) are $a=1$ and $b=1$, and those in Fig. 3(b) are $a=1$ and $b=0.75$. In Figs. 3(a) and 3(b), the volume (solid blue curve) increases quadratically $V_{2D}^{(1)} = \pi r^2$ until $r = a/2$ in Fig. 3(a) and $r = b/2$ in Fig. 3(b) where the solution of Eq. (10) is plotted (solid red curve). For shells with radii equal to or larger than half the diagonal of the detection volume $r \geq \sqrt{a^2 + b^2}/2$, the volume is constant $V = ab$ (black curve). The derivative $\partial V / \partial I$ is plotted along with the volume to check for consistency and correctness. The kernels $K(I, I_0) = \partial V / \partial I$ associated with these three volumes are

$$K_{2D}^{(1)} = K_{2D}^{FV}, \tag{11a}$$

$$K_{2D}^{(2)} = F_{2D}^{FV}\left[\frac{2}{\pi}\arcsin\left(\frac{a}{2r}\right) + \frac{2}{\pi}\arcsin\left(\frac{b}{2r}\right) - 1\right], \tag{11b}$$

$$K_{2D}^{(3)} = 0. \tag{11c}$$

When $r$ is equal to or less than the smallest side of the box, the real part of $K_{2D}^{(2)}$ is equal to $K_{2D}^{(1)}$, and when $r$ is greater than the diagonal length of the detection volume $r \geq \sqrt{(a/2)^2 + (b/2)^2}$ the kernel $K_{2D}^{(2)}$ becomes negative, and by setting the negative values to zero, the kernel $K_{2D}^{(2)}$ is equal to $K_{2D}^{(3)}$. These conditions mean that Eq. (12) can be rewritten as a single expression

$$K_{2D}^{RV} = F_{2D}^{FV}\mathbb{R}\left[\frac{2}{\pi}\arcsin\left(\frac{a}{2r}\right) + \frac{2}{\pi}\arcsin\left(\frac{b}{2r}\right) - 1\right]H_C \tag{12}$$

where $H_C = H(I - I_C)$ is the Heaviside step function with $I_C$ as the intensity at which the function $K_{2D}^{RV}$ changes sign between positive and negative. This occurs when the detection volume is completely within an iso-intensity shell and its derivative is therefore zero. For the 3D case, we find the restricted volume by a hybrid analytical-numerical solution, where the last dimension is numerically integrated over, and from this, the kernel $K_{3D}^{RV}$ can be found. In other words, the 3D restricted kernel $K_{3D}^{RV}$ is constructed by adding exact two-dimensional volume slices from Eq. (12) in which each 2D volume has a width of $\Delta z$ determined by the iso-intensity shells. This can symbolically be written as

$$K_{3D}^{RV} = \sum K_{2D}^{RV} \Delta z . \qquad (13)$$

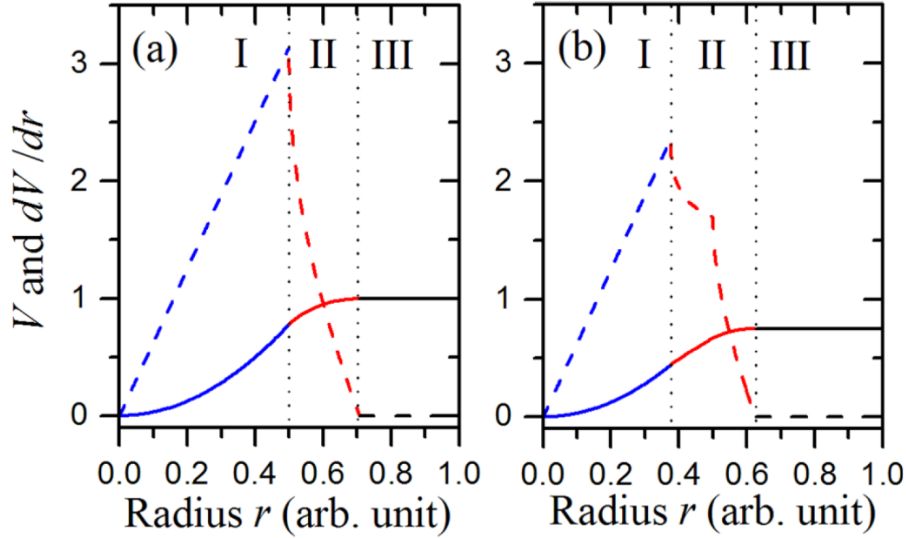

FIG. 3. Detection volume $V(r)$ as a function of radius (solid curve), and derivative of the volume $\partial V / \partial I$ (dashed curve). (a) Volume and its derivative with $a = b = 1$, and (b) volume and derivative with $a = 1$ and $b = 0.75$. Vertical dotted lines separate regions corresponding to different types of restricted volumes.

Equation (13) is one of the main results of this work and gives the volumetric weighting factors for a detection scheme with a restricted volume. For simplicity, we have shown Eq. (13) as a Riemann sum; however, in our calculations, we employed Simpson's rule and equidistant steps in the $z$-direction

## V. RESULTS

Figure 4 shows the result of calculations using Eq. (13) for different detection volumes. In Fig. 4(a), the detection volume has the dimensions $a = 10w_0$, $b = 10w_0$, and $c = 10z_0$. These data (dashed blue curve) were plotted along with the exact result (solid green curve) from Eq. (7c), and the two curves lie on top of one another (for the plot limits shown) demonstrating the accuracy of the calculation. At the peak intensity $I = I_0$, the kernel goes to zero, as it must for all possible slit geometries. If $K_{3D}^{FV}$ did not go to zero here, then the calculations would result in incorrect interpretation of further analysis. In Fig. 4(b), the slit values have been taken to be $a = 10w_0$, $b = 10w_0$, and $c = z_0$. It can be seen that the $K_{3D}^{FV}$ (green) and $K_{3D}^{RV_z}$ (solid red curve) curves cross at an intensity of $I = 0.8I_0$. Here, the calculated kernel $K_{3D}^{RV}$ (dashed blue curve) is plotted along with $K_{3D}^{FV}$ and $K_{3D}^{RV_z}$ and follows $K_{3D}^{FV}$ until an intensity of $I = 0.8I_0$, afterward following $K_{3D}^{RV_z}$ (red curve) for the lower intensities. In Fig. 4(c), the dimensions of the detection volume are $a = w_0$, $b = w_0$, and $c = w_0$. This situation is similar to that shown in Fig. 4(b), except that the calculated kernel deviates from $K_{3D}^{RV_z}$ at an intensity of $I = 0.61I_0$ and goes to zero at $I = 0.36I_0$. Finally, in Fig. 4(d), the slit dimensions are $a = w_0$, $b = 1.5w_0$, and $c = z_0$. This is similar to the data shown in Fig. 4(c), except that the curve after $I = 0.6I_0$ peaks again around $I = 0.32I_0$ and then goes to zero at $I = 0.22I_0$. This analysis shows that the results are in excellent agreement with the known analytical results of Eqs. 7(b) and 7(c) and provides assurance of similar accuracy for the remaining portion of the curves. The calculated kernel $K_{3D}^{RV}$ can be further quantified by determining critical intensities at which the kernel changes geometries as previously discussed. This is the topic of Section VI.

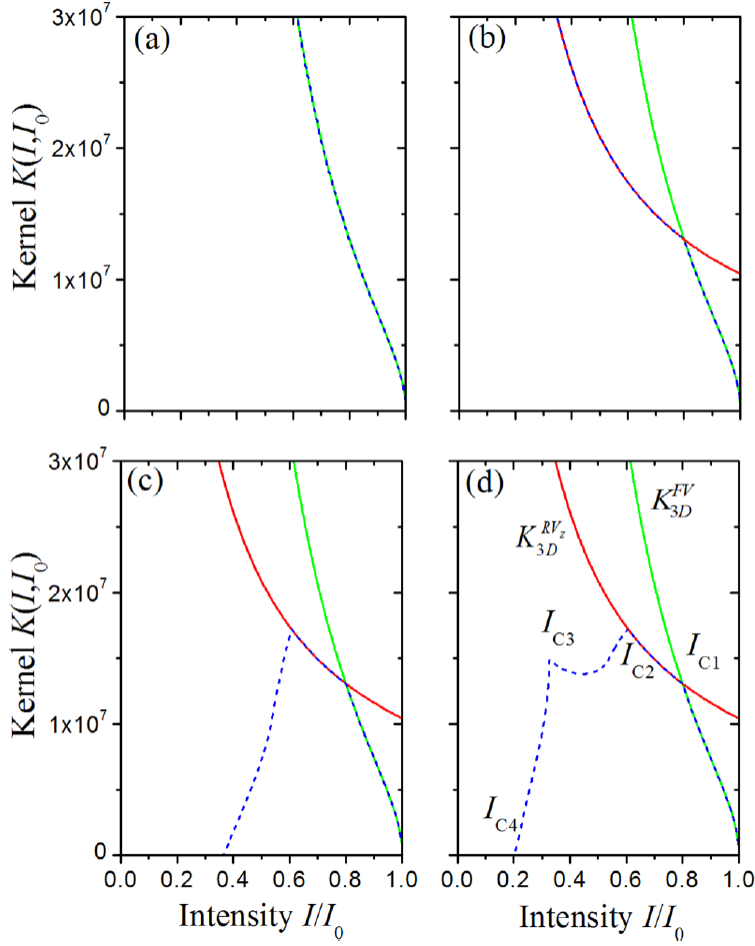

FIG. 4. Theoretically calculated kernels for restricted focal geometries $K_{2D}^{RV}$. (a) Detection volume dimensions $a = b = 10w_0$ and $c = 10z_0$; (b) $a = b = 10w_0$ and $c = z_0$; (c) $a = b = w_0$ and $c = z_0$; and (d) $a = w_0$, $b = 1.5w_0$ and $c = z_0$. In (a), the dimensions of the detection volume are larger than the iso-intensity shells and $K_{3D}^{RV} = K_{3D}^{FV}$ for the plot limits shown. Clipping the iso-intensity shells in the z-direction, $K_{3D}^{RV}$, within a certain range, will be equal to $K_{3D}^{RV_z}$. For intensities lower than $I_{C2}$, the kernel $K_{3D}^{RV}$ can take different forms depending on the values of $a$ and $b$.

## VI. CRITICAL INTENSITIES

In Fig. 4(d), four critical intensities are shown. The dimensions used for the calculated kernel were $a = w_0$, $b = 1.5w_0$, and $c = z_0$. The first critical intensity occurs when the kernel changes from $K_{3D}^{FV}$ to $K_{3D}^{RV_z}$. The kernel $K_{3D}^{FV}$ came about from integration over all space (no restriction) and $K_{3D}^{RV_z}$ was calculated by restricting integration along the z-axis. This suggests that the change in geometry of the kernel is due to clipping of the iso-intensity shells along the z-direction. To quantify this, the local on-axis intensity $I_{0L} = I_0 w_0^2 / w^2$ is calculated at the position $z = c/2$. The intensity found in this way is $I_{C1} = I_0 / \left(1 + c^2 / 4z_0^2\right)$, and for the detection volume used in Fig. 4(d), it is $I_{C1} = 0.80 I_0$. This is the intensity in which the kernel $K_{3D}^{RV}$ changes from that of Eq. (7b) to (7c) due to clipping of the iso-intensity shells in the z direction.

The next critical intensities occur when the largest radius $r(z)$ of a shell meets the transverse dimension of the detection volume. The critical intensity at $0.6 I_0$ occurs because the largest radius of the iso-intensity shell is equal to half the dimension of the smallest transverse length (in this case $a$) of the detection volume. Plugging $r = a/2$ and $z = 0$ into Eq. (2), the critical intensity is found to be $I_{C2} = I_0 \exp(-a^2 / 2w_0^2) = 0.61 I_0$. The same analysis can be applied to the largest transverse slit dimension $b = 1.5 w_0$ and gives an intensity of $I_{C3} = I_0 \exp(-b^2 / 2w_0^2) = 0.32 I_0$. The largest transverse radius $r(z)$ occurs when the derivative is equal to zero, $dr/dz = 0$. In this way, the largest radius occurs at the positions $z = 0$ and $z = \pm\sqrt{I_0 e^{-1} - 1}$. This last expression is real for intensities $I \leq I_0 e^{-1}$, and the isointensity shells take the peanut shapes shown in Fig. 6. When $I \geq I_0 e^{-1}$, the term under the radical is less than zero, and all iso-intensity shells have an elliptical shape with a maximum at $z = 0$. The boundary between the peanut and elliptically shaped iso-intensity shells is shown by the dotted contour in Fig. 5.

Finally, at the lower intensities, isointensity shells can be found such that the detection volume is completely enclosed within it. The highest intensity for which this situation occurs is when the radius of the shell is at the corner of the detection volume at $z = c/2$. To find this intensity, the bracketed expression in Eq. (12) is set to zero, which is motivated by the fact that this results in $K_{3D}^{RV} = 0$. This procedure gives the intensity

$$I_{C4} = I_0 \frac{1}{1 + c^2/4z_0^2} \exp\left(-\frac{1}{2w_0^2}\frac{a^2 + b^2}{1 + c^2/4z_0^2}\right), \tag{14}$$

which for our example gives $I_{C4} = 0.31 I_0$.

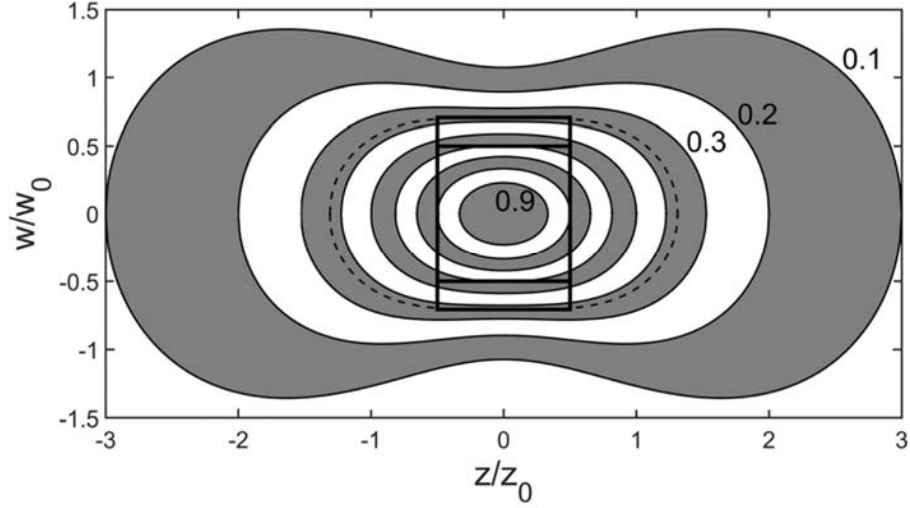

FIG. 5. Isointensity contours and detection volume. The black lines are isointensity contours with heights ranging from $0.1I_0$ to $0.9I_0$ in increments of $0.1I_0$. These contours are either elliptically shaped with a single maximum at $z=0$, or peanut-shaped with two maxima $z = \pm\sqrt{I_0 e^{-1}/I - 1}$. The intensity that separates the elliptical shapes from the peanut shapes is $I = I_0 e^{-1}$ and is shown by the dotted line. The small box represents a detection volume with sides $a = b = w_0$, and $c = z_0$. The larger box is the same as the small box but rotated by 45° and projected onto the $yz$-plane.

In conclusion of this section, while one of the goals of this work was to analyze the origin of the dips in the measured xenon data, our previous work [10] suggested that if a kernel could be found for a restricted focal geometry, then it may be possible to remove residual spatial averaging. That method employed a power series expansion of both the experimental yield $S(I_0)$ and probability $P(I)$ data

$$S(I_0) = I_0^{m-1} \sum_k A_k I_0^k \tag{15a}$$

$$P(I) = I^{m-1} \sum_k \frac{A_k}{G_k} I^k \tag{15b}$$

Here, $k$ is the expansion index, $m$ can be taken to be the lowest positive integer value that results in a solution to the Volterra equation, Eq. (1) (for 1D and 2D geometries $m = 1$ and for 3D geometries $m = 3$),

$A_k$ are expansion coefficients of the yield $S(I_0)$, and the $G_k$ are geometric factors determined by the integrated kernel,

$$G_k \propto \int_0^1 K_{ND}(\xi)\xi^{k+m-1}d\xi. \tag{16}$$

In Ref. 10, the kernels $K(\xi)$ were given for the 1D, 2D, and 3D cases; however, in practice, the kernels are all 3D, and for the kernels given in Eqs. (7b) and (7c), the geometric factors are

$$G_k^{FV} = \pi z_0 w_0^2 N \frac{\sqrt{\pi}}{3}\left[\frac{\Gamma(k+m-5/2)}{2\Gamma(k+m-1)} + \frac{\Gamma(k+m-3/2)}{2\Gamma(k+m)}\right] \tag{17a}$$

$$G_k^{FV_z} = \pi z_0 w_0^2 N \frac{1}{3}\left(\frac{c^2}{z_0^2}+3\right)\frac{1}{k+m-1} \tag{17b}$$

where $N$ is the particle density and $\Gamma$ is the gamma function. To perform the deconvolution, the spatially averaged experimental data $S(I_0)$ are expanded in a power series according to Eq. (15a) to obtain the expansion coefficients $A_k$. The geometric factors $G_k$ are then determined using Eq. (16) with the appropriate volumetric weighting factor $K$. Equations (17) are analytical expressions for the $G_k$ in three dimensional geometries. This work, however, has provided a kernel for restricted focal geometries, Eq. (13), and once calculated using the method outlined here, it can be integrated in Eq. (16) to find the geometric factors. We also note that the kernel $K_{3D}^{RV}$ may contain parts from $K_{3D}^{FV}$ and $K_{3D}^{RV_z}$, which have known solution to Eq. (16). Once the geometric factors are determined, they can be used along with the $A_k$ in Eq. (15b) to recover the ionization probability $P(I)$ in a power series expansion using the new expansion coefficients $A_k/G_k$.

## VII. COMPARISON WITH EXPERIMENTAL DATA

In this section, we investigate the effects of spatial averaging on measured ion yields curves. To do this, we calculated ionization rates $W$ using ADK and PPT theories to generate ionization curves of Xe. Figures 6(a) and 6(b) show the results of the simulated ion yields of xenon up to the seventh charge state. The simulated yields have only a contribution from sequential ionization and show agreement with the data only near saturation. In Fig. 6(a), the measured $Xe^{2+}$ deviates from the calculated yield at an intensity of $\sim 10^{14}\,\text{W/cm}^2$. This intensity corresponds with the saturation intensity of $Xe^{1+}$ and for this reason the $Xe^{1+}$ yields have been shifted up by a factor of 3 times. For comparison with the experimental yields, the simulated yields were spatially averaged using slit dimensions of $a = 12\,\mu m$, $b = \infty$ in the transverse dimension and $c = 400\,\mu m$ along the beam propagation direction. The simulated averaged yields decrease following saturation but do not increase afterwards as seen in the experiment data.

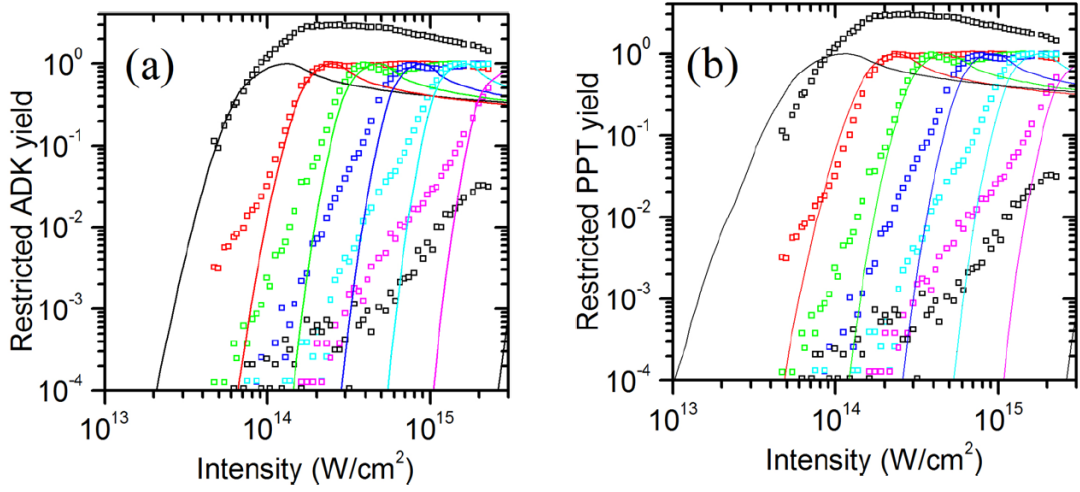

FIG. 6. Ionization yields of Xe$^{n+}$ ($n = 1-7$). (a) Ionization probabilities calculated using ADK theory (solid curves) and measured ionization yields using a detection volume of dimensions $a = 12\,\mu m$ by $b = \infty$ by $c = 400\,\mu m$ (squares). (b) Same as (a) except the probability was calculated using PPT theory.

For better comparison with the data, nonsequential ionization probabilities were simulated by weighting ADK rates of previous charge states according to

$$\frac{dN_n}{dt} = \sum_{j=0}^{n-1} \alpha_{j,n} W_{j,n}^{ADK} N_j - \sum_{k=n+1} \alpha_{n,k} W_{n,k}^{ADK} N_n \qquad (18)$$

Here, $N_n$ is the population of the nth charge state, i.e., $N_0$ is the population of the neutral atoms, the $W_{i,j}$ are the sequential ionization rates determined by ADK, and the $\alpha_{i,j}$ are the nonsequential ionization coefficients [23]. The indices on $\alpha_{i,j}$ and $W_{i,j}$ read from the $i^{th}$ charge state to the $j^{th}$ charge state. For the neutral charge state, the first sum in Eq. (18) is zero, and the rate equation for the neutral population has the analytical solution $N_0 = \exp(-\sum W_{0,k} t)$. Higher order charge state populations were found by numeric integration. The nonsequential ionization coefficients were found by fitting the simulated yields to the experimentally obtained data. Their values are shown in Table I. When all coefficients are zero, except those that are unity, the rate equations lead to sequential ionization yields. The nonsequential yields are shown in Fig. 7(a) and can be contrasted with those in Fig. 6, in which the simulated yields were calculated in a sequential manner. Since the experimental data are relatively flat following saturation, we spatially averaged the simulated probabilities using Eq. (7a) to achieve a better fit. The simulated probabilities were then spatially averaged using the dimensions of the detection volume used in our experiment. These simulated data are plotted in Fig. 7(b). As in the case when ADK and PPT theories were used, a decrease following saturation can be seen but with no increase afterwards. Different slit geometries were investigated; however, complete agreement with the data in Fig. 1 could not be found. Comparison of the results with experiment suggests that the dips in the yield curves may arise from residual spatial averaging.

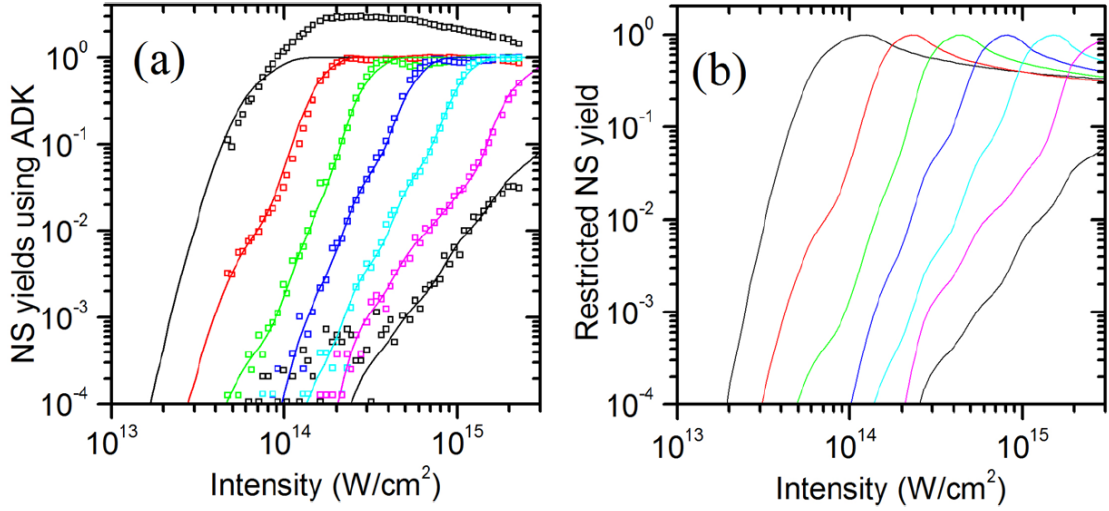

FIG. 7. Spatial averaged nonsequential ionization yields of $Xe^{n+}$ ($n=1-7$). (a) Simulated nonsequential yields spatially averaged using a detection volume with dimensions $a=\infty$ by $b=\infty$ by $c=400\mu m$. (b) Simulated nonsequential yields spatially averaged using our experimental detection volume with dimension of $a=12\mu m$ by $b=\infty$ by $c=400\mu m$. The simulated nonsequential portion of a particular charge state was found by weighting the ADK rates used for previous charge states (see text).

Finally, we attempted to reconstruct the ionization probabilities using Eqs. (15)–(17) with the kernel derived in Eq. (13). The level of noise in the experimental data was such that this technique failed to retrieve the probability; however, we developed a new technique in which the ionization probability at a particular intensity $I_0$ is found from an initial guess. This initial guess was then spatially averaged using Eqs. (1) and (13), and the difference between the simulated signal $Y(I_0)$ and the actual experimental signal $S(I_0)$ was calculated. Based on this difference, a new guess for $P(I_0)$ was made. This procedure was repeated until a minimum was reached between the experimental signal and the simulated signal according to

$$S(I_0) - C\int_0^{I_0} K(I,I_0)P(I)dI \leq \beta S(I_0) \tag{19}$$

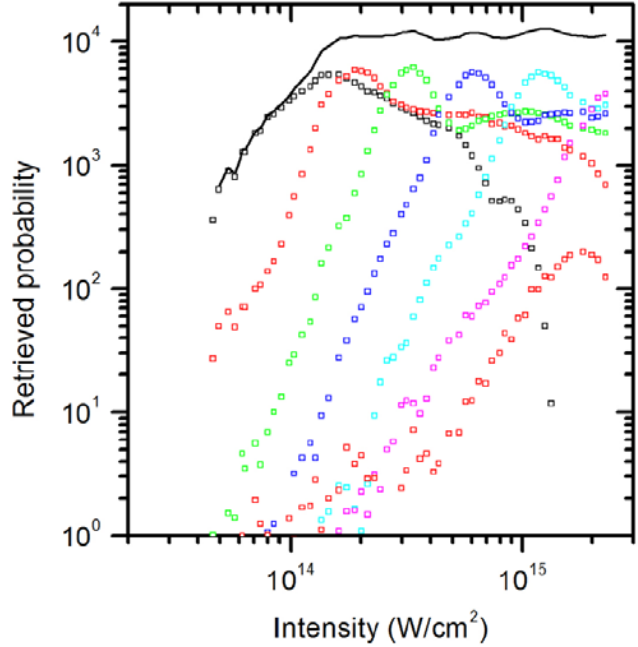

FIG. 8. Retrieved ionization probabilities of Xe charge states. Each charge state is denoted as in Fig. 1. The solid black curve is the sum of all charged states probabilities. For charge states that have reached saturation, a decrease in the probability is observed. The retrieved data show that the probabilities do not continuously decrease as expected by the continuous increase in probability of the following charge state but "stabilize" for some range of intensities before decreasing.

Here, the second term on the left is the simulated signal $Y(I_0)$ with C as a proportionality constant in Eq. (1) and $\beta$ is a closeness parameter. In our simulations, $\beta$ was taken to be equal to $10^{-4}$ of the experimental value $S(I_0)$. As an important note, because $P(I)$ is raised to the power of unity, the solution is unique. This means that when $P(I)$ increases or decreases, the simulated yield increases or decreases, respectively. Figure 8 shows the results of this new deconvolution procedure. The retrieved Xe charge states are as indicated in Fig. 1. The solid black curve is the sum of all charge state probabilities. For charge states up to $Xe^{5+}$, a decrease in probability is seen following saturation as expected from a sequential ionization process. The initial decrease in their probabilities is followed by a region where the probabilities "stabilize" before further decreasing. This decrease and "suppression" in the retrieved ionization probabilities result in the dips observed in the experimental yield curves. The origin of this phenomenon is not known to us, but the structure partially survives the masking effect of spatial averaging in the restricted focal geometry. The integration kernel in Eq. (13) assumes a Gaussian focal geometry, and as a result, the deconvolution is strongly dependent on the form of the focal geometry. While the retrieved ionization probabilities may suffer from imprecise knowledge of the focal intensity profile, we are convinced that the results are consistent with the measured data for two reasons. First, in a 2D detection scheme such as ISS, the yield curves will always be monotonically increasing functions. The fact that we observe a decrease following saturation indicates that our detection volume is more restricted than that of a 2D detection volume. Second, any increase in the yields following a decrease can only occur if the probability stabilizes or increases. The structures in the yield curves are therefore consistent with the retrieved probabilities.

## VIII. CONCLUSIONS

Highly ionized xenon atoms were measured as a function of intensity up to the seventh charge state. Sequential and nonsequential ionization processes were observed in the yields curves. A dip structure was observed in some of the yield curves following saturation where the following charge state has a significant yield. To investigate the dips, we derived for the first time the integration kernel $K_{3D}^{RV}$ for restricted focal geometries. These kernels were used to spatially average simulated ion yields using ADK and PPT theories, and the results partially suggested that the dips may be due to residual spatial averaging. The inclusion of nonsequential multiple ionization had little effect on the spatially averaged curves following saturation. Retrieved probabilities curves found using a new deconvolution technique and the restricted kernel indicate that the dips may be due to a genuine photophysical process.

## ACKNOWLEDGEMENTS

This work was funded by the Robert A. Welch Foundation, Grant No. A1546, and the Qatar Foundation under Grant No. NPRP 5-994-1-172.